\documentstyle[aps,preprint]{revtex}
\begin{document}
\bibliographystyle{prsty}

\draft 

%\preprint{LPS}

\title{Coexisting Pulses in a Model for Binary-Mixture Convection}

\author{Hermann Riecke$^1$ and Wouter-Jan Rappel$^2$\\
$^1$Department of Engineering Sciences and Applied Mathematics,
Northwestern University, Evanston, IL 60208\\
$^2$Department of Physics, Northeastern University,
111 Dana Research Center,
Boston, MA 02115}
\maketitle 

\begin{abstract}
We address the striking coexistence of localized waves (`pulses') 
of different lengths  
which was observed in recent experiments and full numerical 
simulations of binary-mixture convection. Using a set of extended Ginzburg-Landau 
equations, we 
show that this multiplicity finds a natural explanation in terms of 
the competition of two distinct, physical localization mechanisms; 
one arises from dispersion and the other from a concentration mode. 
 This competition is absent in the standard Ginzburg-Landau equation. It may also
be relevant in other waves coupled to a large-scale field.
\end{abstract}

%\pacs{05.45.+b, 05.90.+m, 47.20. Tg}

The rich dynamics of convection in binary mixtures has been studied extensively in
recent years (e.g. \cite{CrHo93}). This system has served as an important paradigm of 
 extended dynamical systems which support nonlinear dissipative waves.
  Perhaps the most striking phenomenon observed is the localization of traveling waves into
`pulses'; in certain parameter regimes convection does not
fill the entire experimental cell but only a small part of it 
\cite{MoFi87,HeAh87,KoBe88,NiAh90,BeKo90}. 
These pulses have raised
various puzzling theoretical questions regarding their localization mechanism and their
propagation velocity. 

Most recently, experimental investigations \cite{Ko94} and
numerical simulations of the full Navier-Stokes equations  \cite{BaLu94}
have revealed in addition a surprising multiplicity of pulses
of different lengths. 
This coexistence of different pulses at the same experimental parameters
has no natural explanation within the standard
complex Ginzburg-Landau equations which are commonly used for the description of 
weakly nonlinear waves. This draws into question the previously established theoretical
 understanding of the localization mechanism of the waves, which 
is based on their dispersive behavior. 
In this Letter we show that a natural understanding
of the experimental as well as the numerical results
is provided by an extension of the Ginzburg-Landau
equations which was introduced previously to explain the anomalously slow drift of the 
pulses \cite{Ri92} and which captures an additional
localization mechanism \cite{HeRi95}.

The essential features of the experiments and the numerical simulations can be 
summarized as follows. In the experiments \cite{Ko94}
pulses have been investigated for three
values of the separation ratio $\Psi$,
 which measures the alcohol concentration of the mixture.
Their length varied from $L=5$ to $L=20$ as measured in vertical gap widths.
For $\Psi =-0.167$ and $\Psi=-0.21$ the pulse length was found to be unique
for any given Rayleigh number and all observed pulses were stable. For $\Psi=-0.127$,
however, the pulse length was found to be not unique over a range of Rayleigh numbers, 
and depending on initial conditions pulses of two different lengths were observed.
While the shorter pulse was stable, the longer pulse was unstable and could only be
found by employing a suitable servo control. Strikingly, the stability
of long pulses seems to be related to their drift velocity:
the {\it stable}, long pulses ($\Psi =-0.167$ and $\Psi=-0.21$) 
drift {\it backward}, i.e. the drift velocity is opposite to
the phase velocity of the waves inside the pulse, 
while the {\it unstable}, long pulses drift {\it forward} ($\Psi=-0.127$).

In the numerical work the full Navier-Stokes equations were simulated
in a quasi-two-dimensional geometry \cite{BaLu91,BaLu94}. 
Stable pulses of a shape similar to that of the experimental
ones were found. As in the experiment, multiple pulse solutions were found to coexist
for the same Rayleigh number (for $\Psi=-0.25$). In contrast to the pulses observed in
experiments, however, all numerical pulses travel forward  (for $\Psi=-0.08$ as
well as $\Psi=0.25$) and {\it both} coexisting pulses were found
to be {\it stable}. 

Previous analytical investigations aimed at explaining the pulses have centered
around the complex Ginzburg-Landau equation (CGL). Investigation of
 two different limits of the CGL showed that dispersion can provide a localization 
mechanism for pulses; while for strong dispersion (i.e. large imaginary parts of 
the coefficients)
the pulses correspond to perturbed soliton solutions of the nonlinear Schr\"odinger
equation \cite{ThFa88,Pi87} they can be considered as a bound pair
of fronts for weak dispersion \cite{HaJa90,MaNe90}. In either case up to two 
solutions exist for given parameters, one
stable and one unstable. However, it is always the {\it large} pulse which is {\it stable}
 and the {\it small} one which is {\it unstable}.  
A similar result was found in the CGL with 
additional gradient terms \cite{LeRaunpub}. 
The unstable solution constitutes 
the `critical droplet' which separates the stable pulse solution from the conductive 
state.
Thus, neither the long, unstable pulses observed experimentally nor the coexistence of stable
pulses of different lengths found in the numerical simulations can be understood within the
CGL. In addition, within the CGL the stability of pulses
is independent of the propagation direction since the group velocity can be scaled away.
 
Here we show that the experimental and the numerical results can be understood quite
naturally within the same set of extended Ginzburg-Landau equations 
(ECGL) that was introduced previously in the context of the anomalously
slow drift of the traveling-wave pulses \cite{Ri92}. We show that the multiplicity and
stability of the experimental and the numerical pulses can be seen to arise 
from the competition of two different localization mechanisms: one due
to dispersion as in the CGL and one due to an additional concentration mode specific
to the ECGL \cite{HeRi95}. 

The extended Ginzburg-Landau equations describe the evolution of the complex convective 
amplitude $A$ coupled to a real concentration mode $C$, which was introduced
to describe the slow mass diffusion in liquids \cite{Ri92}. The relevance of such a
large-scale concentration mode had been identified in the 
full numerical calculations of the Navier-Stokes equations \cite{BaLu91,BaLu94}.
In this paper we focus on a minimal model obtained from the 
equations derived in \cite{Ri92,Ri92a} which contains dispersion and 
the effect of the concentration mode on the local growth rate of the convective mode,
\begin{eqnarray}
\partial_T A+s \partial_X A&=&d \partial^2_X A+(a+fC) A+c A |A|^2 +p|A|^4A, \label{e:ecglA}\\
\partial_T C&=&\delta \partial^2_X C -\alpha C+h_2 \partial_X |A|^2. \label{e:ecglC} 
\end{eqnarray}
The coefficients in (\ref{e:ecglA}), except for the group velocity, are in general complex, 
whereas those in (\ref{e:ecglC}) are real; they
are functions of the system parameters such as the Lewis number, 
which characterizes
the ratio of mass to heat diffusion, and the separation 
ratio $\Psi$, and are given explicitly in 
\cite{Ri92a} for the case of free-slip-permeable boundary conditions. 
Since it is known
that the top and bottom boundary conditions have a strong effect on the values of the 
coefficients we take the values given in \cite{Ri92a} only as an indication of the
behavior of the coefficients with realistic boundary conditions. Thus, 
we take $h_2 f_r$ to be positive ($f \equiv f_r+if_i$), i.e. 
 the advection of alcohol by the traveling convection rolls lead to a reduction of
 the local buoyancy of the fluid ahead of the pulse \cite{Ri92a}.
Such a large-scale current has been identified 
in the full numerical simulations of \cite{BaLu94}. In addition, we follow \cite{Ri92a}
and assume $h_2 f_r$ to 
increase with decreasing $\Psi$. Physically, this can be 
understood  by the fact that the phase velocity of the waves increases with 
decreasing $\Psi$ which enhances the large-scale current. 
For simplicity, we neglect the effect of the
concentration mode on the local frequency of the wave, $f_i=0$. In the framework of
a perturbed nonlinear Schr\"odinger equation it has been discussed in \cite{Ri95}.

In view of the experimental and the numerical results \cite{Ko94,BaLu94}
 we solve (\ref{e:ecglA},\ref{e:ecglC}) numerically and study, in particular,
 the dependence of the length of pulse solutions of (\ref{e:ecglA},\ref{e:ecglC})
on $a_r$, which is proportional to the Rayleigh number. 
Motivated by the dependence of the coupling strength $h_2f_r$ on $\Psi$ 
we focus first on two cases: a case of strong coupling to the concentration mode 
 and a case of weaker coupling.  

\begin{figure}
\begin{picture}(420,300)(0,0)
\put(0,-30) {\includegraphics{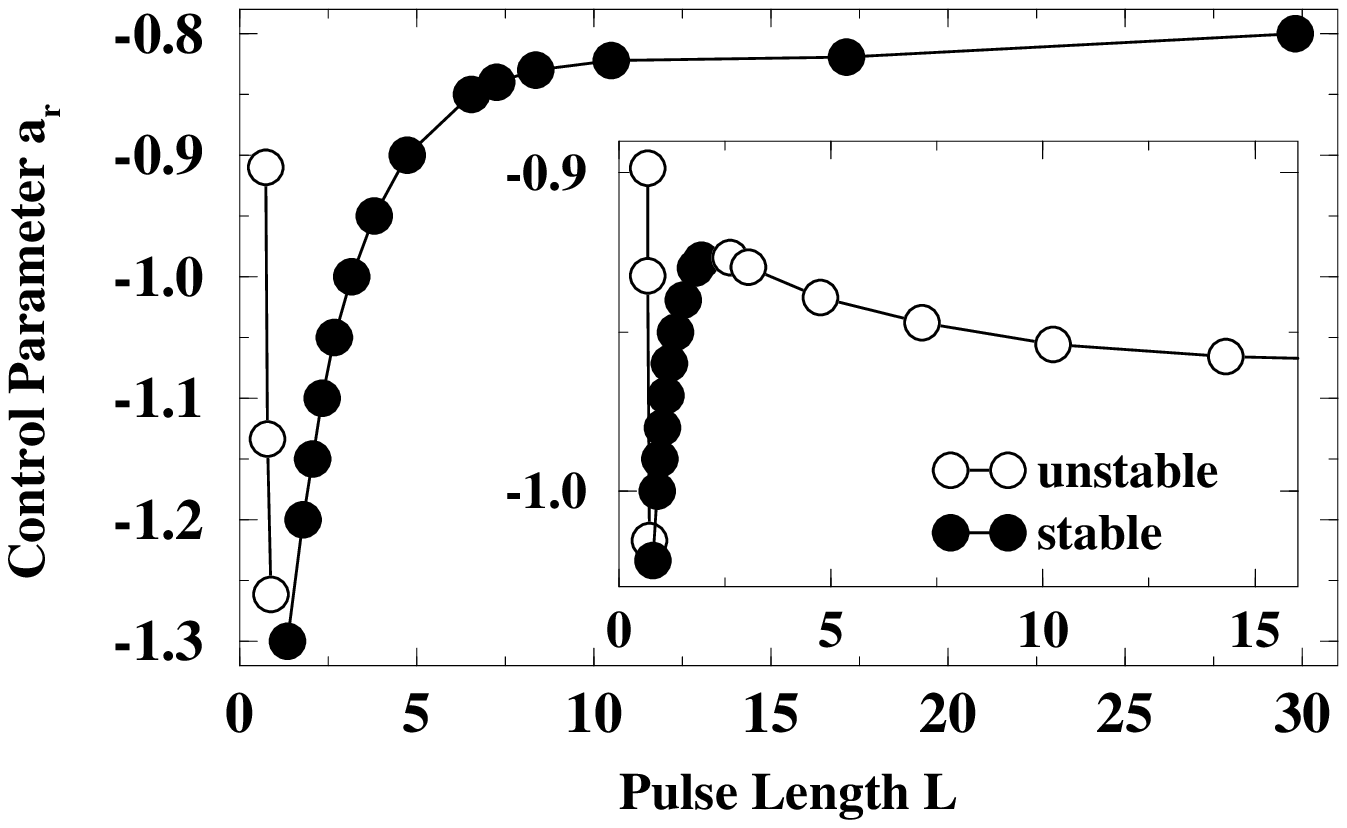}}
\end{picture}
\caption{Pulse length {\it vs.} growth rate $a_r$
for $d=0.01$, $c=2.45-0.5i$, $p=-1$, 
$\delta=0.009$, $\alpha=0.02$, $s=0.1$ and $f=0.1$.
For $h_2=0.3$ 
the pulses travel backward for all values of 
$a_r$ whereas for $h_2=0.05$ (inset) 
they travel forward. 
\protect{\label{f:LarC1}}
}
\end{figure}

Fig.\ref{f:LarC1} gives the pulse length $L$ as a function
of $a_r$ for two values of the coupling $h_2$; 
$h_2=0.3$ and $h_2=0.05$ (inset). The
remaining parameters are indicated in the caption.
For $h_2=0.3$ there is a branch of solutions (solid circles) on which
the pulse length increases
monotonically with  $a_r$
 and the pulses are stable
all along the branch. The branch of unstable solutions
(open circles) separates the stable pulse solution 
from the conductive state. These solutions constitute therefore the critical
droplet. Due to the strong reduction of the
buoyancy ahead of the pulse the stable and the unstable pulses travel backward.

With decreasing coupling the pulse velocity increases and for $h_2=0.05$  (inset)
the pulses travel forward for all $a_r$. 
As in the case $h=0.3$, there is a branch corresponding to the very short, unstable 
critical droplets (open circles) and a branch of short, stable pulses (solid circles).
Now, however, the latter branch turns around for larger distances rendering
the long pulses unstable.

The simplicity of the ECGL allows the identification of the mechanisms that lead to the
stability results shown in fig.\ref{f:LarC1} by considering the pulses as consisting of 
two interacting fronts\footnote{In the present
paper we are not addressing the interaction between pulses. As shown in \cite{Ri95a}, 
the central features of the experimental results on the interaction \cite{Ko91a} 
can also be understood within (\ref{e:ecglA},\ref{e:ecglC}). Striking differences
are found as compared to the interaction without the concentration mode \cite{BrDe89a}.}.
The interaction of such fronts has been studied previously in two different limits.
The contribution from dispersion alone has been investigated within the CGL in the limit
of widely separated fronts ($L$ large) \cite{MaNe90,HaJa90}. The interaction {\it via} the 
concentration mode alone has been studied within the ECGL
in the limit of fronts with narrow width $\xi$ \cite{HeRi95}. 
Based on these two different analyses we expect 
 that the essential features of the combination of both effects can be modelled by
an evolution equation for the length $L$ of the pulse which 
has the form\footnote{More precisely, one
would expect coupled equations for the velocities of the leading and of the trailing front
which both depend also on the distance between them \cite{HeRi95}.} (see also \cite{HeRi95})
\begin{equation}
\partial_T L= k_0 (a_r -a_r^e) -k_1 e^{-L/\xi}+\frac{k_2}{L}-\frac{k_3}{v}
e^{-\alpha L/|v|}. \label{e:dLdt}
\end{equation}
The first term describes the invasion of the conductive state by the convective state
for $a_r$ larger than the equilibrium value $a_r^e$ ($k_0>0$). 
The second term represents the 
attractive interaction ($k_1>0$) between fronts which arises already within the real
Ginzburg-Landau equation. The third term gives the contribution to the interaction from
dispersion. 
In the absence of the concentration mode, stable pulses exist only if
$k_2>0$. 
The last term captures the interaction {\it via} the concentration mode $C$. It
 decays over a length which depends on the 
damping $\alpha$ of $C$ and the velocity $v$ of the pulse. 
In the limit considered in \cite{HeRi95} $k_3=\sqrt{12}h_2  f_r$. For the regime
in question $k_3$ is therefore positive \cite{Ri92a,BaLu94}.
Most importantly, the sign of the interaction 
depends on the direction of propagation of the pulse. 

\begin{figure}
\begin{picture}(420,300)(0,0)
\put(0,-30) {\includegraphics{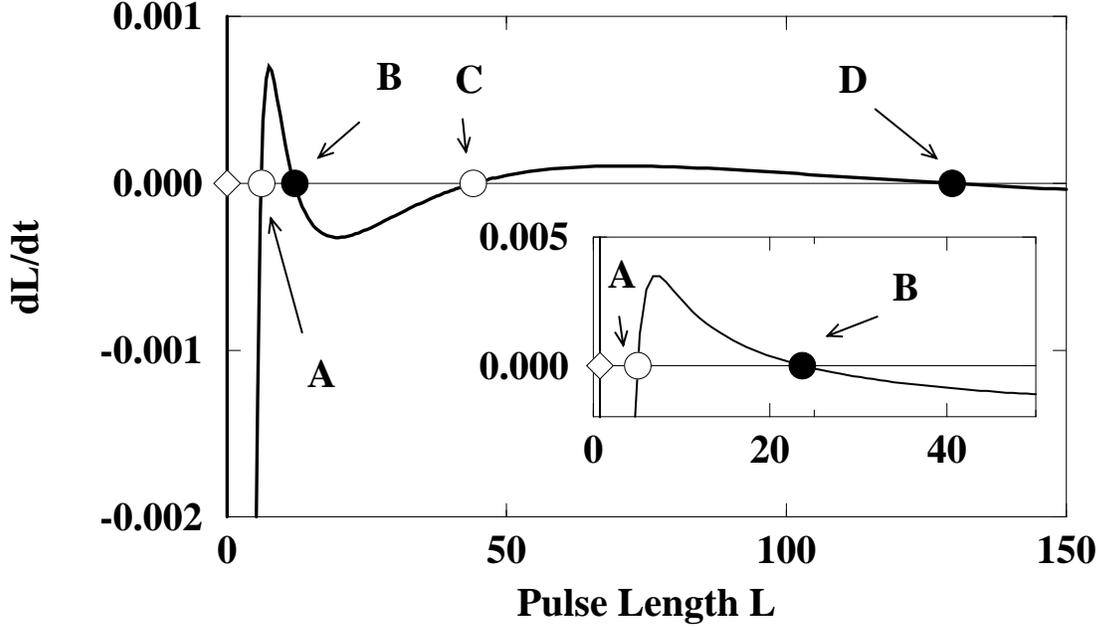}}
\end{picture}
\caption{Steady-state solutions of 
(\protect{\ref{e:dLdt}}): $dL/dt=-0.0003-e^{-x}+0.04/x-0.0055 e^{-0.05 x}$. 
The solid circles 
indicate stable pulse solutions the open circles unstable ones. The diamond
gives the unphysical solution with \protect{$L=O(\xi)\equiv O(1)$}. The inset shows
steady-state solutions of (\protect{\ref{e:dLdt}}) for $v<0$:
$dL/dt=-0.002-e^{-x}+0.04/x+0.001 e^{-0.05 x}$.
\protect{\label{f:dLdt}}
}
\end{figure}

To interpret the numerical results shown in fig.\ref{f:LarC1}
we plot in fig.\ref{f:dLdt} the right-hand-side of (\ref{e:dLdt}) for
$v<0$ (inset) and for $v>0$. For $v<0$ (\ref{e:dLdt}) allows
 three solutions. The longest pulse is stable (full circle) and the shorter one unstable 
(open circle). The smallest pulse (open diamond) has $L=O(\xi)$ and 
falls outside the range of validity of (3). This corresponds to the strong-coupling
case $h_2=0.3$ in fig.\ref{f:LarC1}. For forward traveling pulses, $v>0$, 
(\ref{e:dLdt}) has five solutions: two stable, two unstable, and one unphysical
solution. This corresponds to the weak-coupling case ($h_2=0.05$) if the stable
solution in fig.\ref{f:LarC1} is identified with pulse B and the unstable ones with
A and C. Note, that increasing $a_r$ raises the curve in fig.\ref{f:dLdt} and leads
to a merging of solutions B and C as in fig.\ref{f:LarC1}. For these parameter values
no stable long pulses corresponding
to solution D were found in the numerical simulations of (\ref{e:dLdt}) (see below).

The results shown in fig.\ref{f:LarC1} 
agree qualitatively very well with the experiments of Kolodner 
(cf. figs.32,33 in \cite{Ko94}).
There also, short pulses are stable for all $\Psi$ investigated, 
independent of their direction of propagation.
Long pulses, on the other hand, are only stable for strongly negative $\Psi$ (i.e. strong 
coupling) for which the pulses travel backward.
For positive velocity (larger $\Psi$) the long pulses are unstable. Most
importantly, the comparison of fig.\ref{f:LarC1} and fig.\ref{f:dLdt}
 allows the identification of the relevant mechanisms
in the different experimental regimes. For short pulses the dispersive interaction
 dominates and the localization mechanism is essentially as given by the CGL.
For the longer pulses, however, 
the interaction $via$ the concentration mode is stronger. Therefore, the pulse
 stability is closely related to the propagation velocity and long  pulses
are unstable for weakly negative $\Psi$. The cross-over between the two mechanisms can 
be seen in the experimental data for strongly negative $\Psi$ where the Rayleigh-number
dependence of the length shows a clear break at intermediate lengths.

As mentioned earlier, in the numerical simulations of the Navier-Stokes equations
multiple pulse solutions have been obtained as well \cite{BaLu94}. 
In contrast to the experiments, however, 
the short as well as the long pulse were found to be stable. This difference may be 
due to the fact that in the numerical simulations a quasi-two-dimensional geometry 
was employed while the experiments were done in a narrow 
channel\footnote{The difference in geometry 
will also be reflected in the values of the coefficients of (\ref{e:ecglA},\ref{e:ecglC}).}. 

\begin{figure}
\begin{picture}(420,300)(0,0)
\put(0,-30) {\includegraphics{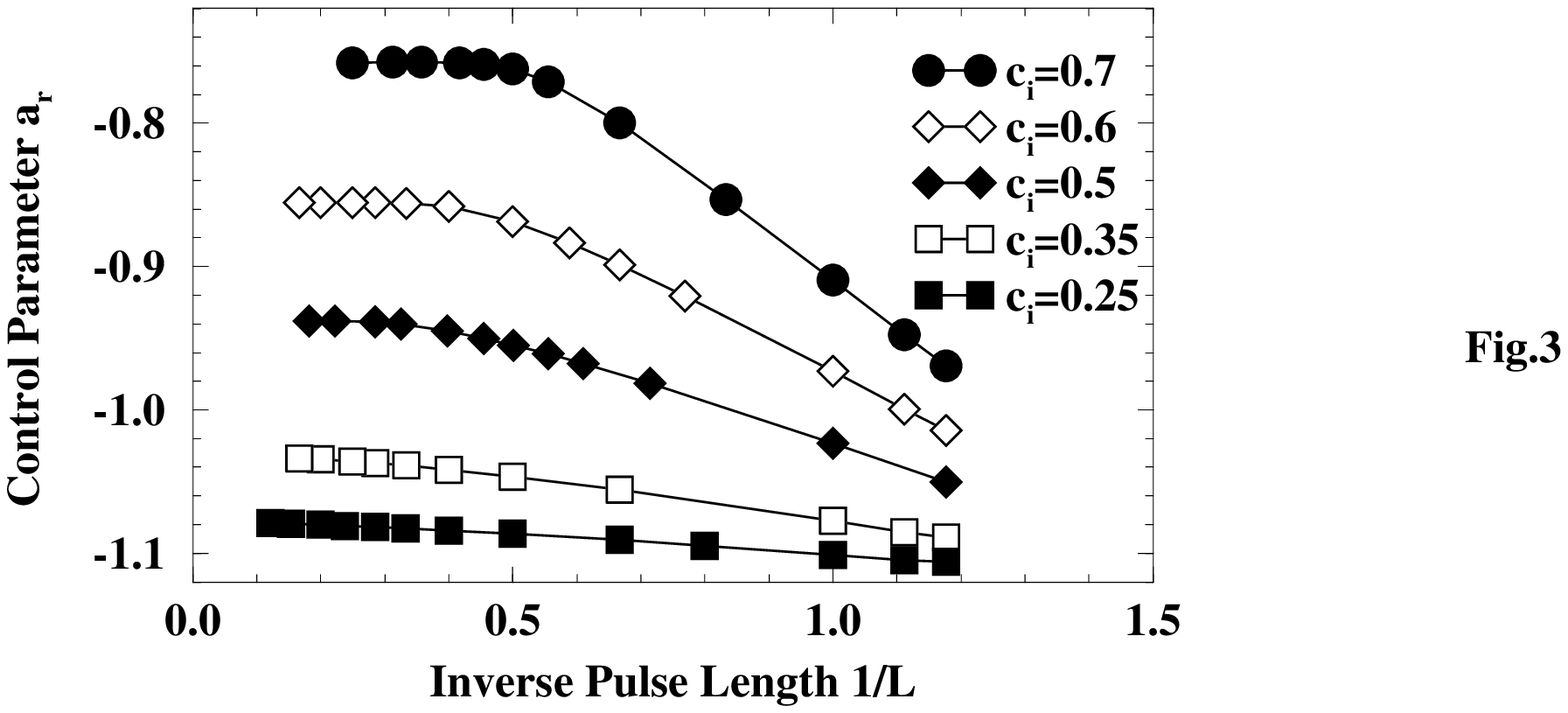}}
\end{picture}
\caption{Pulse length $L$ {\it vs.} control
parameter $a_r$ in the absence of the concentration 
mode ($f=0$) for $d=0.01$, $c_r=2.45$, $p=-1$.
The power-law behavior extends only up to a maximal length 
$L_{max}$ which decreases with increasing dispersion. 
\protect{\label{f:Larf0}}
}
\end{figure}

Are the ECGL (\ref{e:ecglA},\ref{e:ecglC}) able to account also for this
coexistence of {\it stable} pulses of different length? 
The evolution equation (\ref{e:dLdt}), which is conjectured based on (\ref{e:ecglA},\ref{e:ecglC}), suggests that quite generally there should be a second 
 stable pulse (the long pulse D) since the power-law behavior of the 
dispersive term dominates the 
exponentially decaying interaction {\it via } the
concentration mode for very large  $L$. 
 A more detailed analysis of the dispersive interaction 
shows, however, that the power-law behavior persists only up to 
some maximal length $L_{max}$
which decreases with increasing dispersion \cite{MaNe90}.
To study the dispersive interaction over the full range of $L$ 
we investigate the ECGL in the absence of the
 concentration mode (i.e. the usual CGL).
Fig.\ref{f:Larf0} shows  $a_r$  as a function of the inverse length of the pulse, $1/L$, for
various different values of the dispersion $c_i$.
Clearly the power-law behavior is  
only found up to some maximal length $L_{max}$
which decreases with increasing dispersion. Beyond this length the dispersive
interaction decays extremely rapidly. Therefore, if in the full ECGL the concentration 
mode should dominate for lengths up to $L_{max}$ and beyond,
dispersion will not stabilize long pulses.

\begin{figure}
\begin{picture}(420,300)(0,0)
\put(0,-30) {\includegraphics{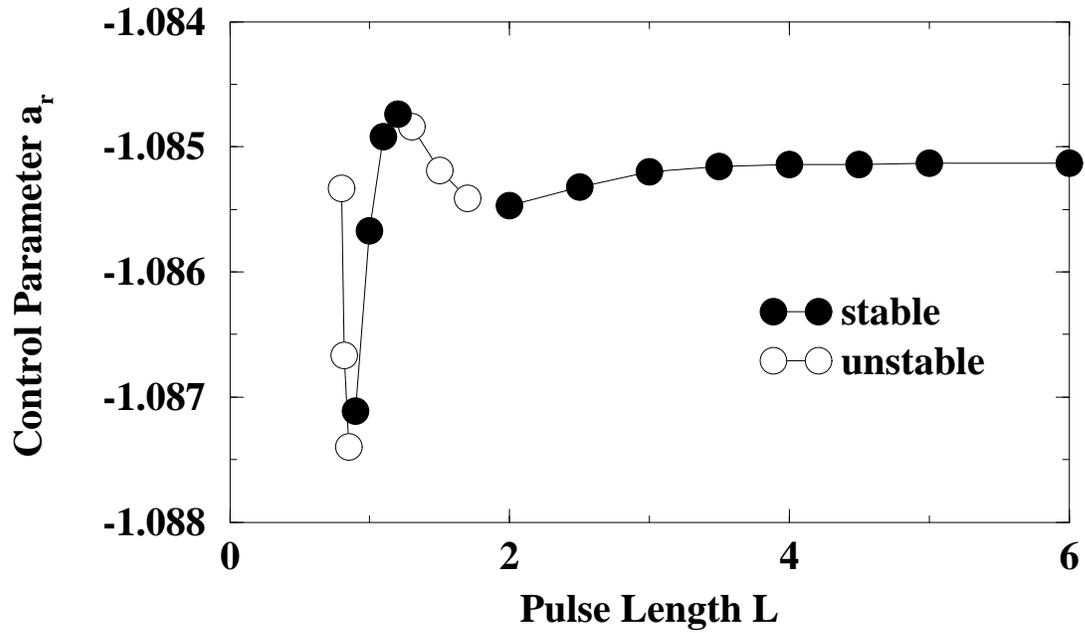}}
\end{picture}
\caption{Pulse length {\it vs.} growth rate $a_r$ 
for a rapidly decaying concentration  mode 
($d=0.01$, $c=2.45-0.25i$, $p=-1$,
$\delta=0.03$, $\alpha=0.5$, $s=0.1$, $h_2=0.05$
and $f=0.45$).
For large $L$ the dispersion becomes dominant 
again and restabilizes the pulses leading to the coexistence of stable pulses.
\protect{\label{f:LarC2}}
}
\end{figure}

Thus, the coexistence of short as well as long stable pulses obtained
 in \cite{BaLu94} is only expected to arise if
the contribution from the concentration mode decays sufficiently fast.
Fig.\ref{f:LarC2} shows the results of simulations of (\ref{e:ecglA},\ref{e:ecglC})
with increased damping $\alpha$ as well as larger coupling strength $f_r$. 
Indeed, two unstable and two stable pulses are found, in
agreement with the evolution equation (3). This gives a natural explanation of the
coexistence of stable pulses obtained in the full numerical simulations  \cite{BaLu94}
in terms of an alternating dominance of two different localization mechanisms. 

The robustness of the localization mechanism due to 
the concentration mode suggests
that the extended Ginzburg-Landau equations may also allow
localized chaotic pulses. Temporally chaotic localized waves have been found previously
in the complex Ginzburg-Landau equation \cite{DeBr94} as well as in the
extended Ginzburg-Landau equation in the absence of dispersion \cite{HeRiunpub}.
In these cases it is essentially the length of the pulse which varies chaotically; the
traveling
wave state within the pulse is regular. In the presence of dispersion the extended Ginzburg-Landau
equation possibly allows waves which are localized by the concentration mode
and which at the same time are  
Benjamin-Feir unstable and exhibit phase chaos. 

In conclusion, we have shown that the extended Ginzburg-Landau equation can account for
the essential features of the localized waves found in experimental and numerical 
investigations of binary-mixture convection; the ordering of stable and unstable
pulses as well as the coexistence of stable pulses of different lengths is seen to be a 
natural consequence of the competition between the interaction between fronts due
to dispersion and due to the concentration mode. The latter depends 
sensitively on the direction of propagation of the pulse. Thus concentration-dominated
long pulses are stable  only if they travel backward in agreement with
the experimental results \cite{Ko94}.  If the concentration mode decays
sufficiently fast in space two coexisting stable pulses of different lengths can 
arise as found in numerical simulations of the Navier-Stokes equations \cite{BaLu94}.
 The results demonstrate that the 
concentration field plays a crucial role in the stability and
the dynamics of the pulses. Its essential features are captured at least qualitatively
by the extension of the Ginzburg-Landau equations discussed in this paper. It is worth 
mentioning that the extended Ginzburg-Landau equations apply also to
interfacial waves in Poiseuille flow \cite{ReRe93} and waves coupled to a large-scale
field in general. They may also be relevant for electroconvection in
nematic liquid crystals \cite{TrKrpriv}.  

{\bf Acknowledgments.} H.R. gratefully acknowledges stimulating discussions with
H. Herrero. This work was supported by  the EEC programme
`Human Capital and Mobility' (WJR) and by DOE and NSF through grants
DE-FG02-92ER14303 and DMS-9304397 (HR). 

\bibliography{/home/hermann/.index/journal} 

\newpage

\newpage

\newpage

\end{document}